\documentclass[twocolumn,nofootinbib,amsmath,amssymb,superscriptaddress,
a4paper]{revtex4}

\usepackage{graphicx}
\usepackage{dcolumn}
\usepackage{bm}

\begin{document}

%
\title{Non-CKM induced flavor violation in ``minimal'' SUSY SU(5)
 models
}

\author{Francesca Borzumati}
\affiliation{ICTP, Strada Costiera 11, 34014 Trieste, Italy}
\affiliation{Department of Physics, National Central University, 
             Chung-Li 32054, Taiwan}
\author{Satoshi Mishima}
\affiliation{School of Natural Sciences, Institute for Advanced Studies,
             Princeton, NJ 08540, USA}
\author{Toshifumi Yamashita}
 \thanks{Talk presented by T.~Yamashita at the Workshop
 {\it ``Flavour in the era of the LHC''}, CERN, October 9-11 2006, 
 and the {\it ``4th International Workshop on the CKM unitarity
 triangle''}, Nagoya, Japan, December 12-16 2006}
\affiliation{SISSA and INFN, Via Beirut 4, I-34014 Trieste, Italy}

\begin{abstract}
Patterns of flavor violation induced by neutrino Yukawa couplings are
 discussed in realistic ``minimal'' SUSY SU(5) models, obtained by
 adding nonrenormalizable operators to the minimal one, in order to
 fix the fermion spectrum and suppress proton decay. 
Results are presented for the three possible implementations of the 
 seesaw mechanisms, i.e. of Type~I,~II and~III.
\end{abstract}

\maketitle
\section{Introduction and motivations}
\vspace{-0.2truecm} 
%
Supersymmetry~(SUSY) is still one of the most interesting possibilities
 to solve the hierarchy problem of the standard model~(SM) of particle
 physics.
A solution to this problem without excessive tuning requires that the
 massive parameters breaking SUSY softly are around the TeV scale,
 which, for simplicity, is hereafter identified with the electroweak
 scale $M_{\rm weak}$.

Irrespectively of the extensions needed to solve the hierarchy problem, 
 the leptonic sector requires also an extension of the originally
 proposed SM structure of only three left-handed SU(2) doublets, in 
 order to accommodate neutrino masses.
One way to proceed is to introduce SM singlets, or right-handed
 neutrinos~(RHNs), which can couple to the leptonic doublets with
 Yukawa couplings of ${\cal O}(1)$ if their Majorana masses are
 superheavy. 
This is the conventional and well-known seesaw mechanism, which enjoys
 immense popularity because of its elegance, but which is difficult to
 test experimentally.
It is therefore very important to search for signals that can give
 information on the existence of the heavy particles realizing this
 mechanism.
An obvious magnifying glass for them could be precisely their large
 Yukawa couplings to the left-handed leptons, $Y_\nu$, and the large
 leptonic mixing angles in the MNS matrix.
Indeed, these couplings can affect sizably the renormalization
 group~(RG) flow of the soft SUSY-breaking parameters for the
 sleptons~\cite{LFV<-NuYUKAWA} from the cutoff scale, at which the
 breaking of SUSY is mediated to the visible sector, $M_{\rm cut}$,
 down to the seesaw scale $M_{\rm seesaw}$.
They lead to non-vanishing off-diagonal elements of the
 charged-slepton mass matrix at $M_{\rm weak}$, or 
 lepton-flavor violations~(LFVs) in the left-left sector of this
 matrix, $\tilde m^2_{e_{LL}}$. 
The existence of intrinsic flavor violations in the slepton mass
 parameters at $M_{\rm cut}$, however, could completely obscure 
 the effects of the RHN interactions through RG equations~(RGEs).
Thus, we restrict ourselves to considering models with 
 flavor-blind SUSY breaking and mediation of this breaking.

If in addition, we embed these SUSY models in a grand unified
 theory~(GUT), the RHNs interact with these large Yukawa couplings also
 with the right-handed down quarks, which are the SU(5) partners of the
 doublet leptons.
Hence, as pointed out by Moroi~\cite{QFV<-NuYUKAWA}, these interactions
 can affect also the massive soft parameters of the down-squark sector,
 generating quark-flavor violations~(QFVs) in the scalar sector
 different from those induced by the quark Yukawa couplings. 
In particular, in the superCKM basis for quark superfields, the scalar
 QFVs due to the RHNs are in the right-right sector of the down-squark
 mass matrix, $\tilde m^2_{d_{RR}}$, whereas those induced by the
 quark Yukawa couplings in non-GUT setups are in the left-left
 one~\cite{HKR}.
(GUT phases also appear when identifying the SM fields among the 
 components of the SU(5) multiplets. Here, we neglect them altogether,
 postponing the discussion of their effect to a later occasion.)
Thus, it has been argued that, in SUSY SU(5) models with RHNs and
 flavor-blind soft massive parameters at $M_{\rm cut}$, scalar LFVs
 and QFVs at $M_{\rm weak}$ are related to each
 other~\cite{QFV<-NuYUKAWA} in a simple way.

The minimal model, however, is not realistic: it predicts a
 too rapid proton decay and the wrong relation between the down-quark
 mass matrix and the charged-lepton's one. 
New physics beyond that of the minimal SUSY SU(5) model is needed to
 cure these problems and it is easy to imagine that such additional
 degrees of freedom can modify even drastically the simple
 relations between LFVs and QFVs of Ref.~\cite{QFV<-NuYUKAWA}, and 
 of many successive works. 
We refer to these relations as Moroi's predictions.

As is well known, one way to fix the incorrect fermion spectrum 
 consists in the introduction of nonrenormalizable
 operators~(NROs), suppressed by $1/M_{\rm cut}$. 
The effects on flavor violation of only one such NRO of dimension-five
 (sufficient for the purpose) were studied in
 Ref.~\cite{flNROinSU5}. 
They amount to introducing some arbitrariness in the choice of the
 flavor rotations of the SM fields when they are embedded in the SU(5)
 multiplets. 
This is expressed by the appearance of two additional unitary matrices
 (other than the RGE-evolved CKM and MNS ones), with arbitrary mixings
 among the first two generations, but with smaller ones among the
 third and the second/first generations.
In the parameter space of mixings/phases opened up by the introduction
 of this NRO, there is however still a region in which these unitary
 matrices of additional mixings reduce to the unit matrix. 
In this region, the pattern of intergenerational sfermion mixings
 remains unchanged with respect to that obtained without NROs, i.e.
 Moroi's predictions for flavor transitions, then, can be kept as
 viable.
The authors of  Ref.~\cite{flNROinSU5}, however, did not discuss the
 problem of a too-large decay rate of the proton, induced by the
 exchange of colored Higgs fields.
One way to suppress it, compatible with their analysis, is to assume
 that there exist other NROs, also suppressed by $1/M_{\rm cut}$, that
 are baryon-number violating and that cancel (up to experimentally 
 tolerable remnants) the colored-Higgs-fields induced operators
 responsible for proton decay. 
These, indeed, are dimensionally suppressed by the inverse mass of the
 colored Higgs fields, supposed to be larger than $1/M_{\rm cut}$, but
 are also further suppressed by coefficients depending on small Yukawa
 couplings and small CKM mixing angles.
Hence, the cancellation is expected to be possible, and the model of
 Ref.~\cite{flNROinSU5} can be made realistic 
 with some tuning (this is in addition to the intrinsic tuning 
 required in this model for the doublet-triplet mass splitting).
It remains, however, to be checked whether the parameter space of
 additional mixings/phases relevant for flavor transitions remains
 unchanged for all values of $\tan\beta$, once this cancellation is
 enforced.

As outlined in Ref.~\cite{DESYpeople}, a technically different way to
 suppress proton decay becomes possible, if the number of NROs employed
 to fix the fermion spectrum is enlarged. 
It was shown by the authors of Ref.~\cite{DESYpeople} that even only
 the addition of four NROs of dimension five is sufficient to introduce
 enough SU(5)-breaking effects to disentangle the Yukawa couplings
 contributing to the coefficient of the effective operators responsible
 for proton decay from the couplings giving rise to fermion masses and
 mixings. 
At the expenses of some tuning, then, it is possible to make these
 effects large enough to reduce the rate for proton decay below 
 experimental limits, even for colored Higgs fields with mass of 
 ${\cal O}(M_{\rm GUT})$, where $M_{\rm GUT}$ is the so-called GUT scale. 
An enlargement of the number of NROs allows even more freedom to achieve 
 this suppression~\cite{HABAN}.

Motivated by these considerations, we try to go one step further and
 study the relations between LFVs and QFVs in realistic ``minimal''
 SUSY SU(5) models~\cite{BMY}, with up to an infinite number of NROs
 added to the minimal SU(5) structure. 
These models share with the truly minimal one the fact that the Higgs
 sector solely responsible for the breaking of the SU(5) and the SM
 symmetries is given by the three Higgs multiplets ${5}_H$,
 ${\bar5}_H$, and ${24}_H$, with superpotential:
\begin{eqnarray}
W_H  = 5_H (M_5 + \lambda_5 24_H )\bar{5}_H 
 +\!{\textstyle \frac{1}{2}} M_{24}24_H^2 
 +\!{\textstyle \frac{1}{6}}\lambda_{24}24_H^3. \quad
\label{eq:WH}
\end{eqnarray}
We remind here that $5_H$ and $\bar{5}_H$ contain the two weak Higgs
 doublets $H_u$ and $H_d$ of the minimal supersymmetric SM, and two
 color triplets, $H^C_U$ and $H^C_D$, i.e.
  $5_H = \{H^C_U,H_u\}$ and $\bar{5}_H=\{H^C_D, H_d\}$. 
The ${24}_H$ has among its components $G_H$, $W_H$ and $B_H$, 
 which are adjoint fields of SU(3), SU(2) and U(1), respectively. 
It contains also the vector-like pair $X_H$ and $\bar{X}_H$, with
 $X_H$ ($\bar{X}_H$) a triplet (antitriplet) of SU(3) and a doublet of 
 SU(2).
The SM quark and lepton fields $Q$, $U^c$, $D^c$, $L$, and $E^c$ are
 collected in the two matter multiplets ${10}_M = \{Q,U^c,E^c\}$ and
 ${\bar5}_M =\{D^c,L\}$, with one replica of them for each generation,
 interacting according to
\begin{equation}
W_M =
   {\scriptstyle \sqrt{2}} \,\bar5_M Y^5 10_M \bar5_H 
  -{\textstyle \frac{1}{4}} \, 10_M Y^{10} 10_M 5_H. 
\label{eq:WM}
\end{equation}
Apart from the obvious extensions needed to accommodate neutrino masses,
 these models differ from the truly minimal one for the addition of
 NROs. 
We treat them in as much generality as it is possible, for example by
 including practically all classes of those needed for the fermion
 spectrum, of all dimensions, since we find that the problem is
 actually technically manageable. 
This, however, does not exclude that some of these coefficients are
 accidentally vanishing. 
In this sense, if enough NROs explicitly violating baryon number are
 introduced to suppressed proton decay in the way outlined above, also
 the model of Ref.~\cite{flNROinSU5}, with only one NRO used to fix 
 the fermion spectrum, becomes part of this class of models.
We refrain from studying here the flavor predictions for this 
 modification of the model of Ref.~\cite{flNROinSU5}, but we restrict
 ourselves to models in which the suppression of the proton-decay rate
 is achieved with a procedure of the type outlined in
 Ref.~\cite{DESYpeople}.
Interestingly, this procedure is predictive.
Since it involves a specific flavor ansatz for the Yukawa couplings
 mediating the proton decay rate, it fixes some of the additional
 mixings obtained in Ref.~\cite{flNROinSU5}: it leaves Moroi's
 predictions for flavor transitions between sfermions in the
 $\bar{5}_M$ representation of SU(5) unchanged, while induces
 modifications for those in the $10_M$ representations~\cite{BMY}.
As for flavor transitions in the $\bar{5}_M$ sector, we try to
 investigate what other type of ultraviolet physics may affect them.
One obvious way to do that is to implement possible different types
 of the seesaw mechanism. 
We review them in Sec.~\ref{seesaw}.  
Another way is to disentangle the cutoff scale from the reduced
 Planck mass, $M_P$, by taking it as an adjustable scale varying from
 $M_P$ and $M_{\rm GUT}$. 
Values of $M_{\rm cut}$ below $M_P$ are for example typical of models
 with gauge mediation of SUSY breaking; they can occur also when the
 ``minimal'' models are embedded in higher-dimensional
 setups~\cite{CUTOFF}. 
We show some results in Sec.~\ref{analysis}, after having specified 
 the value of parameters used in this analysis. 

\begin{figure}
\includegraphics[width=0.47\textwidth]{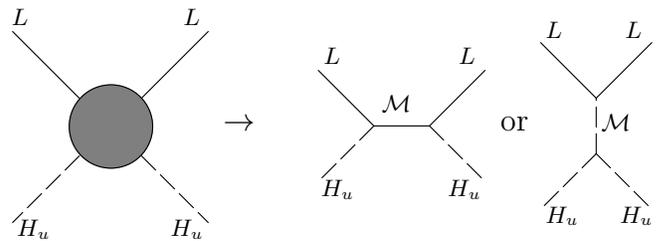}
\put(-240,80){$L$}
\put(-175,80){{{$L$}}}
\put(-238,0){{{$H_u$}}}
\put(-180,0){{{$H_u$}}}
\put(-100,47){{{${\cal M}$}}}
\put(-122,65){$L$}
\put(-71,65){{{$L$}}}
\put(-123,15){{{$H_u$}}}
\put(-75,15){{{$H_u$}}}
\put(-17,40){{{${\cal M}$}}}
\put(-38,75){$L$}
\put(-8,75){{{$L$}}}
\put(-38,5){{{$H_u$}}}
\put(-10,5){{{$H_u$}}}
\put(-160,40){{\large{$\to$}}}
\put(-55,40){{\large or}}
\caption{\label{fig:seesaw} The seesaw mechanism.}
\end{figure}
%
\section{Seesaw Mechanism}    
\vspace{-0.2truecm} 
\label{seesaw}
%
The seesaw mechanism is a mechanism to generate the effective
 dimension-five operator for neutrino masses, $LH_u LH_u$, by
 integrating out heavy degrees of freedom at the scale
 $M_{\rm seesaw}$. 
It is depicted schematically in Fig.~\ref{fig:seesaw}.
In this figure, a solid (broken) line indicates a fermion (boson) or, 
 in a supersymmetric context, a superfield with an odd (even) 
 $R$-parity.
At the tree level, there are only two diagrams that can give rise to the
 effective operator, one mediated by a solid line and one
 by a broken line.
At first glance, it might seem that the inner line, representing the 
 mediator ${\cal M}$, can be a singlet or triplet of SU(2) in both cases.
In reality, the possibility of the singlet scalar is forbidden by 
 the multiplication rule of SU(2):
$ 2\times2=1_A+3_S$,
 where the indices $A$ and $S$ indicate an anti-symmetric and symmetric
 product, respectively. 
Thus, there are only three types of seesaw mechanism, distinguished by
 the nature of the mediator, which can be an SU(2) 
\begin{center}
\begin{tabular}{lcl}
 singlet fermion & \quad  - \quad & Type I, \\
 triplet scalar  & \quad  - \quad & Type II, \\
 triplet fermion & \quad  - \quad & Type III, 
\end{tabular}
\end{center}
 i.e. the RHNs $N^c$, a triplet Higgs $T$ and what we call 
 matter triplets $W_M$, respectively.

Their interactions with the SU(2) lepton doublets are 
\begin{equation}
 N^c  Y_\nu^{\rm I} L H_u, \quad
 {\textstyle \frac1{\sqrt2}} L Y_\nu^{\rm II} T L,  \quad
 {\scriptstyle \sqrt2} \,H_u W_M Y_\nu^{\rm III} L. 
\label{eq:NUYukawaInteracs}
\end{equation}
Integrating out the mediators and replacing $H_u$ by its vev $v_u$, we
 obtain the effective neutrino mass matrices:
\begin{equation}
 m_\nu = \left\{
 \begin{array}{lcl}
 \left(Y_\nu^{\rm I,III}\right)^T \!
  \displaystyle{\frac{1}{M_{\cal M}^{\rm I,III}}}
 \left(Y_\nu^{\rm I,III}\right) v_u^2   
 \\[1.2ex]
 Y_\nu^{\rm II} \displaystyle{\frac{\lambda_U}{M_{\cal M}^{\rm II}}}
 \,v_u^2 
 \end{array}
 \right.
\label{eq:NUmassmatrix}
\end{equation}
 in the three cases. 
Here $M_{\cal M}^{\rm I,III}$ are mass matrices whereas 
 $M_{\cal M}^{\rm II}$ is a number, and $\lambda_U$ is the coupling of
 $H_uTH_u$.
In Type II, because the mediator has no flavor, the high-energy input
 in the neutrino mass matrix is just a number, i.e. the ratio
 $\lambda_U/M_{\cal M}^{\rm II}$, and the flavor structure of
 $Y_\nu^{\rm II}$ is the same as that of the neutrino mass:
\begin{equation}
 Y_\nu^{\rm II}= \frac{1}{v_u^2}
   V_{\rm MNS}^* \,{\hat m_\nu} V_{\rm MNS}^T 
 \frac{M_{\cal M}^{\rm II}}{\lambda_U},
\label{eq:TypeIIrel}
\end{equation}
where $\hat{m}_\nu$ is the diagonal form of $m_\nu$, and $V_{\rm MNS}$
 is the MNS matrix including here two Majorana phases. 
This is in great contrast with the Type~I and~III, in which the
 mediators carry flavor indices. 
In these cases, the flavor structure of $Y_\nu^{\rm I,III}$ is
 different from that of $m_\nu$ and there is a large number of
 high-energy parameters contributing to the neutrino mass matrix, which
 can be expressed in terms of the three eigenvalues of
 $M_{\cal M}^{\rm I,III}$, $(\hat M_{\cal M}^{\rm I,III})_{ii}$, and
 an arbitrary complex orthogonal matrix $R$~\cite{CASAS}: 
\begin{equation}
 \left(Y_\nu^{\rm I,III}\right)^T  =
   \frac{1}{v_u}
   V_{\rm MNS}^* \sqrt{\hat m_\nu}\,R\sqrt{\hat M_{\cal M}^{\rm I,III}}.
\label{eq:CASASrel}
\end{equation}
Notice also that in these two cases, $m_\nu$ is quadratic in
 $Y_\nu^{\rm I,III}$,
 whereas in the Type II seesaw it is linear in $Y_\nu^{\rm II}$.

When embedded in an SU(5) GUT, the multiplets containing these
 mediators are matter singlets, $N^c$, in the case of the Type~I
 seesaw, a Higgs field in a 15plet, $15_H$, in Type~II, and finally in
 Type~III, adjoint matter fields, $24_M$.
The Yukawa interactions in Eq.~(\ref{eq:NUYukawaInteracs}) become now
\begin{equation}
 -N^c Y_N^{\rm I}\bar{5}_M 5_H,                                  \quad
{\textstyle \frac1{\sqrt2}}\bar{5}_M Y_N^{\rm II} 15_H \bar{5}_M,\quad 
 5_H 24_M Y_N^{\rm III} \bar{5}_M, 
\label{eq:SU5NUYukawaInteracs}
\end{equation}
 which contain many more SM interactions than those listed in 
 Eq.~(\ref{eq:NUYukawaInteracs}).
($Y_N^{\rm I,II,III}$ and $Y_\nu^{\rm I,II,III}$ differ by phase
 factors, as discussed in Ref.~\cite{BMY}.)
As anticipated in the introduction, then, the large off-diagonal
 entries in $Y_\nu$ can affect not only the leptonic sector, but also
 the hadronic one. 
%
\begin{table}
\caption{The SU(5) Yukawa interactions of the seesaw mediators, 
 together with their SM decompositions, and the expected patterns of 
 flavor violations are listed.}
\begin{ruledtabular}
\begin{tabular}{lccc}
                &&&\\[-1.9ex] 
            & Type I           & Type II               & Type III 
\\[1.01ex]\hline\hline&&&\\[-1.9ex]
mediator    & $N^c$            & $15_H$                & $24_M$ 
\\[1.01ex]
interaction & $N^c\bar5_M5_H$  & $\bar5_M15_H\bar5_M$  
                               & $5_H24_M\bar5_M$ 
\\[1.01ex]\hline\hline&&&\\[-1.9ex]
only LFV    
  & \begin{tabular}{c} $N^cLH_u$      \\ -         \end{tabular}      
  & \begin{tabular}{c} $LTL$          \\ -         \end{tabular}    
  & \begin{tabular}{c} $H_uW_ML,\,H_uB_ML$\\ $H_U^CX_ML$ \end{tabular}
\\[1.05ex]\hline&&&\\[-1.9ex]
LFV\,\&\,QFV&    -             & $D^cLQ_{15}$          &    -  
\\[1.01ex]\hline&&&\\[-1.9ex]
only QFV    
  & \begin{tabular}{c}  -             \\ $N^cD^cH_U^C$ \end{tabular}
  & \begin{tabular}{c} $D^cSD^c$      \\ -             \end{tabular}
  & \begin{tabular}{c} $H_u\bar X_M D^c$
                                       \\ $H_U^CG_M D^c,\,H_U^CB_M D^c$ 
                                                       \end{tabular}
\\[1.9ex]\hline\hline&&&\\[-1.5ex]
LFV/QFV & $>1$ & $\sim1$ & $\sim1$ 
\end{tabular}
\end{ruledtabular}
\label{tab:seesaw}
\end{table}
%
%
Indeed, the SM decomposition of the interactions in 
 Eq.~(\ref{eq:SU5NUYukawaInteracs}) is given in Table~\ref{tab:seesaw}.
The SM interactions are accommodated in different lines depending 
 on whether they give rise to off-diagonal terms in the left-left
 sector of the charged-slepton mass matrix, in the right-right
 sector of the down-squark mass matrix, or in both. 
The fields $Q_{15}$ and $S$ in the column ``Type II'' and 
 $B_M$, $X_M$, $\bar{X}_M$ and $G_M$ in the column ``Type III''
 are the SU(5) partners of the Higgs triplet $T$ and
 of the triplet fermion $W_M$, respectively.
It should be noticed here that the colored Higgs field $H_U^C$ decouples
 at $M_{\rm GUT}$, which is at least two orders of magnitude larger than
 $M_{\rm seesaw}$, where $N^c$, $15_H$, and $24_M$ are integrated out.
Therefore, below $M_{\rm GUT}$, only the interactions without $H_U^C$ 
 remain active. 
Thus, in the Type I seesaw, LFVs in the scalar sector are in general 
 larger than QFVs, as the 
 interaction $N^cD^cH_U^C$ decouples earlier than $N^cLH_u$.
In contrast, in the Type II seesaw, LFVs and QFVs are of the same
 order up to sub-leading SU(5)-breaking effects in the RG flows below
 $M_{\rm GUT}$. 
This is simply due to the fact that the full SU(5) interaction
 remains active down to $M_{\rm seesaw}$. 
As for the Type III, because two of the interactions inducing LFVs and
 one of those inducing QFVs survive between $M_{\rm GUT}$ and
 $M_{\rm seesaw}$, the relations between LFVs and QFVs depend on
 group-theoretical factors. 
An explicit calculation shows that their magnitudes are of the same
 order. 
The situation is summarized in the last line of Table~\ref{tab:seesaw}.

\begin{figure}
\includegraphics[width=0.43\textwidth]{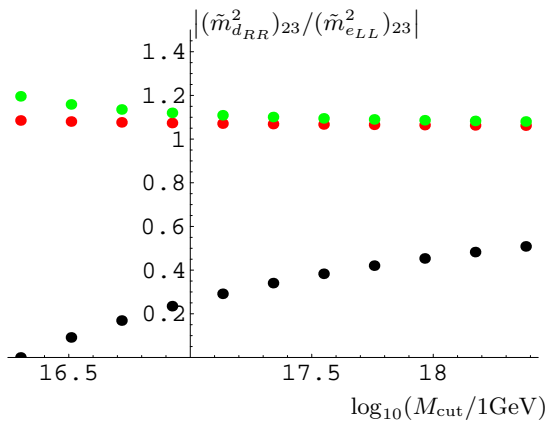}
\put(- 80, -10){{{$\log_{10}(M_{\rm cut}/1{\rm GeV})$}}}
\put(-140, 135){{{$\left|{(\tilde m^2_{d_{RR}})_{23}}
                          /{(\tilde m^2_{e_{LL}})_{23}}\right|$}}}
\caption{The ratio
$|{(\tilde m^2_{d_{RR}})_{23}}/{(\tilde m^2_{e_{LL}})_{23}}|$
 as a function of $M_{\rm cut}$.
The lower line of black dots show the results obtained for the 
 Type~I seesaw, 
 the two upper lines of red (below) and green (above) dots those 
 for the Type~II and~III, respectively.}
\label{fig:DoverL23} 
\end{figure}

\vspace{-0.2truecm} 
\section{Analysis and summary}
\vspace{-0.2truecm} 
\label{analysis}
%
We summarize the choice of parameters made for our analysis.
The cutoff scale $M_{\rm cut}$ is varied from $M_{\rm GUT}$ to 
 $M_P=2.4\times10^{18}{\rm GeV}$. 
Of the four parameters in Eq.~(\ref{eq:WH}), two are needed to 
 fix $M_{\rm GUT}$ and the mass of the colored Higgs fields $H_U^C$
 and $H_D^C$.
We take both these parameters to be $2\times10^{16}{\rm GeV}$. 
This choice is consistent with the unification of gauge couplings
 and with the bounds coming from the proton-decay
 rate~\cite{DESYpeople}. 
One remaining parameter of the four in Eq.~(\ref{eq:WH}) is needed to
 finetune the electroweak scale; the fourth is free. 
We choose this to be $\lambda_{24}$. 
Throughout our analysis we take this to be of ${\cal O}(1)$.
In particular, in the plots that we show here, it is fixed to be
  $1/2$.  
As for the parameters of the Type~II seesaw, we set $\lambda_U=1/2$,
 and $M_{\cal M}^{\rm II} = M_{\rm seesaw}=10^{14}{\rm GeV}$. 
For the Type~I and~III, we take the $R={\bf1}$, and similarly
 $M_{\cal M}^{\rm I,III}=\hat{M}_{\cal M}^{\rm I,III} =
 M_{\rm seesaw}{\bf1}$, with the same  value of $M_{\rm seesaw}$ used
 for the Type~II.
In the light-neutrino sector, we adopt the normal hierarchy of masses. 
The mixing angle $\theta_{13}$ and all three phases of $V_{\rm MNS}$
 are set to zero. 
As for the soft SUSY-breaking parameters, we go beyond flavor blindness
 and assume universality at $M_{\rm cut}$, as usually done in these
 analyses~\cite{QFV<-NuYUKAWA,flNROinSU5}.
We fix the gaugino mass, $M_{1/2}$, the common scalar mass,
 $\tilde{m}_0$, and the common proportionality constant in the
 trilinear couplings, $A_0$, to be $1\,$TeV. 
Finally we take $\tan \beta=10$.

We are now in a position to show some results. 
We solve the RGEs from $M_{\rm cut}$ to $M_{\rm weak}$, reported in
 Ref.~\cite{BMY}, for the entries $(2,3)$ in the mass matrices
 $\tilde m^2_{d_{RR}}$ and $\tilde m^2_{e_{LL}}$, for the three possible
 implementation of the seesaw mechanism, and for different values of
 $M_{\rm cut}$. 
We plot in Fig.~\ref{fig:DoverL23} the absolute value of the ratio of
 these entries as a function of $M_{\rm cut}$. 
The three different lines of dots correspond to the three different
 types of seesaw mechanism. 
As foreseen in Sec.~\ref{seesaw} the mixing (2,3) induced in
 $\tilde m^2_{e_{LL}}$ is larger than that in $\tilde m^2_{d_{RR}}$
 induced by the same neutrino Yukawa coupling in the seesaw of Type~I. 
See lower line of black dots in this figure. 
As also expected, the down-squark mixing decreases when $M_{\rm cut}$
 approaches $M_{\rm GUT}$ as the interval in which this mixing is
 induced becomes shorter.
The two upper lines of red and green dots show the results obtained for
 the seesaw mechanisms of Type~II and~III, in agreement with the
 expectations discussed in Sec.~\ref{seesaw}. 
The results shown in this figure remain pretty much unchanged for
 different choices of the GUT parameters, soft SUSY-breaking parameters,
 and type of neutrino-mass hierarchy chosen. 
They are obtained using a flavor ansatz as in Ref.~\cite{DESYpeople},
 to suppress proton decay, having used an unlimited number of NROs to
 fix the fermion spectrum~\cite{BMY}.
As explained in the introduction, they are consistent with the 
 predictions by Moroi for the seesaw of Type~I, with $M_{\rm cut}=M_P$.
We note, however, that the analysis of Ref.~\cite{flNROinSU5} would give
 results for the ratio of the (2,3) elements of $\tilde m^2_{d_{RR}}$
 and $\tilde m^2_{e_{LL}}$ in general plagued by the uncertainty of
 additional mixings/phases (uncertainty possibly reduced when
 suppressing proton decay in the way outlined in the introduction).
In summary, we conclude this section, with the observation that 
 flavor transitions, do depend, in general, on the detailed
 implementations of NROs used to cure the problem of the minimal 
 SUSY SU(5) model.

\vspace{-0.4truecm} 
\begin{acknowledgments}
\vspace{-0.2truecm} 
\noindent F.~B. is supported in part by the Taiwan NCS grants 
 No~95-2811-M-008-043 and No~95-2112-M-008-001,
 S.~M. by the US DOE grant No.~DE-FG02-90ER40542. 
T.~Y. acknowledges partial financial support from Shizuoka University,
 Japan and thanks the theory group at the NCU,
 Taiwan, where parts of this work were carried out. 
\end{acknowledgments}

\end{document}